\documentstyle[seceq,letter,twocolumn]{jpsj2}

\def\runauthor{Kenji {\sc Goto} {\it et al.}}

\title
{
Pressure-Induced Magnetic Quantum Phase Transition from Gapped Ground State in TlCuCl$_3$
}

\author
{ 
Kenji {\sc Goto}\footnote{E-mail: goto@lee.phys.titech.ac.jp.}, Masashi {\sc Fujisawa}, Toshio {\sc Ono}, Hidekazu {\sc Tanaka}$^{1}$ and Yoshiya {\sc Uwatoko}$^{2}$
}

\inst
{
Department of Physics, Tokyo Institute of Technology, Oh-okayama, Meguro-ku, Tokyo 152-8551\\
$^{1}$Research Center for Low Temperature Physics, Tokyo Institute of Technology, Meguro-ku, Tokyo 152-8551 \\
$^{2}$Institute for Solid State Physics, The University of Tokyo, 
Kashiwanoha, Kashiwa, Chiba 277-8581\\
}

\recdate
{
September 8, 2004}

\abst
{
Magnetization measurements under hydrostatic pressure were performed on an $S=1/2$ coupled spin dimer system TlCuCl$_3$ with a gapped ground state under magnetic field $H$ parallel to the $[2, 0, 1]$ direction. With increasing applied pressure $P$, the gap decreases and closes completely at $P_{\rm c} = 0.42\pm 0.05$ kbar. For $P > P_{\rm c}$, TlCuCl$_3$ undergoes antiferromagnetic ordering. A spin-flop transition was observed at $H_{\rm sf}\simeq 0.7$ T. The spin-flop field is approximately independent of pressure, although the sublattice magnetization increases with pressure. The gap and N\'{e}el temperature are presented as functions of pressure. The occurrence of the pressure-induced quantum phase transition is attributed to the relative enhancement of the interdimer exchange interactions compared with the intradimer exchange interaction.
}

\kword
{
TlCuCl$_3$, spin gap, pressure-induced ordering, quantum phase transition, quantum critical point
}

\begin{document}
\sloppy
\maketitle

A quantum phase transition (QPT) is a phase transition between different quantum ground states induced by a continuous change in interaction constants or applied field \cite{Sachdev}. In the vicinity of the transition point called the quantum critical point (QCP), physical properties are governed by quantum fluctuations. Novel ground states are stabilized by quantum fluctuations and the critical behavior associated with QPT is of current interest in condensed matter physics. In this paper, we report on the magnetic QPT induced by the applied pressure in the coupled spin dimer system TlCuCl$_3$.

TlCuCl$_3$ crystallizes in a monoclinic structure \cite{Takatsu}, which is the same as the structure of KCuCl$_3$ \cite{Willett}. The crystal structure consists of planar dimers of Cu$_2$Cl$_6$, in which Cu$^{2+}$ ions have spin $1/2$. The magnetic ground state is a spin singlet with an excitation gap of ${\it \Delta} /k_{\rm B}=7.5$ K \cite{Takatsu,Shiramura,Oosawa1}. The origin of the spin gap is the antiferromagnetic exchange interaction $J/k_{\rm B}=65.9$ K on the planar dimer Cu$_2$Cl$_6$ \cite{Cavadini1,Oosawa2}. The markedly small gap as compared with the intradimer exchange interaction $J$ is ascribed to strong interdimer exchange interactions with the three-dimensional network.

Because of the small gap, the field-induced magnetic QPT from the gapped spin liquid state to the antiferromagnetic state with transverse-ordered moments can be observed in TlCuCl$_3$ using a conventional superconducting magnet \cite{Oosawa1}. Thus, the static and dynamic properties of the field-induced magnetic QPT have been extensively investigated in various experiments \cite{Tanaka1,Rueegg,Sherman}. The results were clearly described in terms of the Bose-Einstein condensation of spin triplets called magnons or triplons \cite{Sherman,Nikuni,Matsumoto2,Misguich}.

It is considered that the application of hydrostatic pressure enhances interdimer interactions, because the distances between dimers are contracted. Since the spin gap shrinks with increasing the magnitude of interdimer interactions, and, in general, the effect of pressure on exchange interactions is fairly large in chlorides, we can expect the occurence of a pressure-induced magnetic QPT in the pressure range of $P<10$ kbar, which is easily accessible with a high-pressure cramp cell. With this reasoning, we performed magnetization measurements on TlCuCl$_3$ under hydrostatic pressure. 

The magnetizations were measured at temperatures down to 1.8 K under magnetic fields up to 7 T using a SQUID magnetometer (Quantum Design MPMS XL). Pressures up to 10 kbar were applied using a cylindrical high-pressure cramp cell designed for use with the SQUID magnetometer \cite{Uwatoko}. A sample of size $2.5\times 2.5\times 5$ mm$^3$ was set in the cell with its $[2, 0, 1]$ direction parallel to the cylindrical axis. The $[2,0,1]$ direction is parallel to two cleavage planes, $(0, 1, 0)$ and $(1, 0, -2)$. A magnetic field was applied along the $[2, 0, 1]$ direction. As pressure-transmitting fluid, both a mixture of Fluorinert FC70 and FC77, and Daphne oil 7373 were used. The pressure was calibrated using the superconducting transition temperature $T_{\rm c}$ of tin placed in the pressure cell. The diamagnetism of tin was measured at $H=10$ and 50 Oe to determine $T_{\rm c}$ after removing the residual magnetic flux trapped in the superconducting magnet. The accuracy of the pressure is 0.1 kbar for the absolute value and 0.05 kbar for the relative value.

Figure 1 shows the magnetization curves obtained at various pressures at 1.8 K. The data were corrected for the magnetization due to the pressure cell. At ambient pressure, the magnetization is small up to the critical field $H_{\rm g}=5.5$ T due to the spin gap, after which it increases rapidly. The critical field $H_{\rm g}$ is related to the spin gap ${\it \Delta}$ at zero field as $H_{\rm g}={\it \Delta}/g\mu_{\rm B}$ with $g=2.06$ measured by ESR. It is evident that the critical field decreases with increasing pressure. The magnetization of the ground state is proportional to $H-H_{\rm g}$ just above $H_{\rm g}$, and the slope of the magnetization curve decreases with decreasing $H_{\rm g}$ \cite{Matsumoto2}. For this reason and the finite temperature effect, the bend anomaly at $H_{\rm g}$ observed at 1.8 K becomes smeared with increasing pressure. Therefore, there is a certain amount of error in the determination of $H_{\rm g}$. For the evaluation of the critical fields $H_{\rm g}$ for $P < 0.39$ kbar, we used two fitting functions for the magnetizations, the linear function ($K_0+K_1H$) for $H < H_{\rm g}$ and the quadratic function ($K_0'+K_1'H+K_2'H^2$) for $H > H_{\rm g}$, because the magnetization curve is expressed using a convex function of $H$ above $H_{\rm g}$ \cite{Shiramura,Matsumoto2}. The critical fields indicated by arrows in Fig. 1(a) were evaluated from a field at which the two fitting functions cross.
\begin{figure}
\begin{center}
\includegraphics[width=7.8cm,clip]{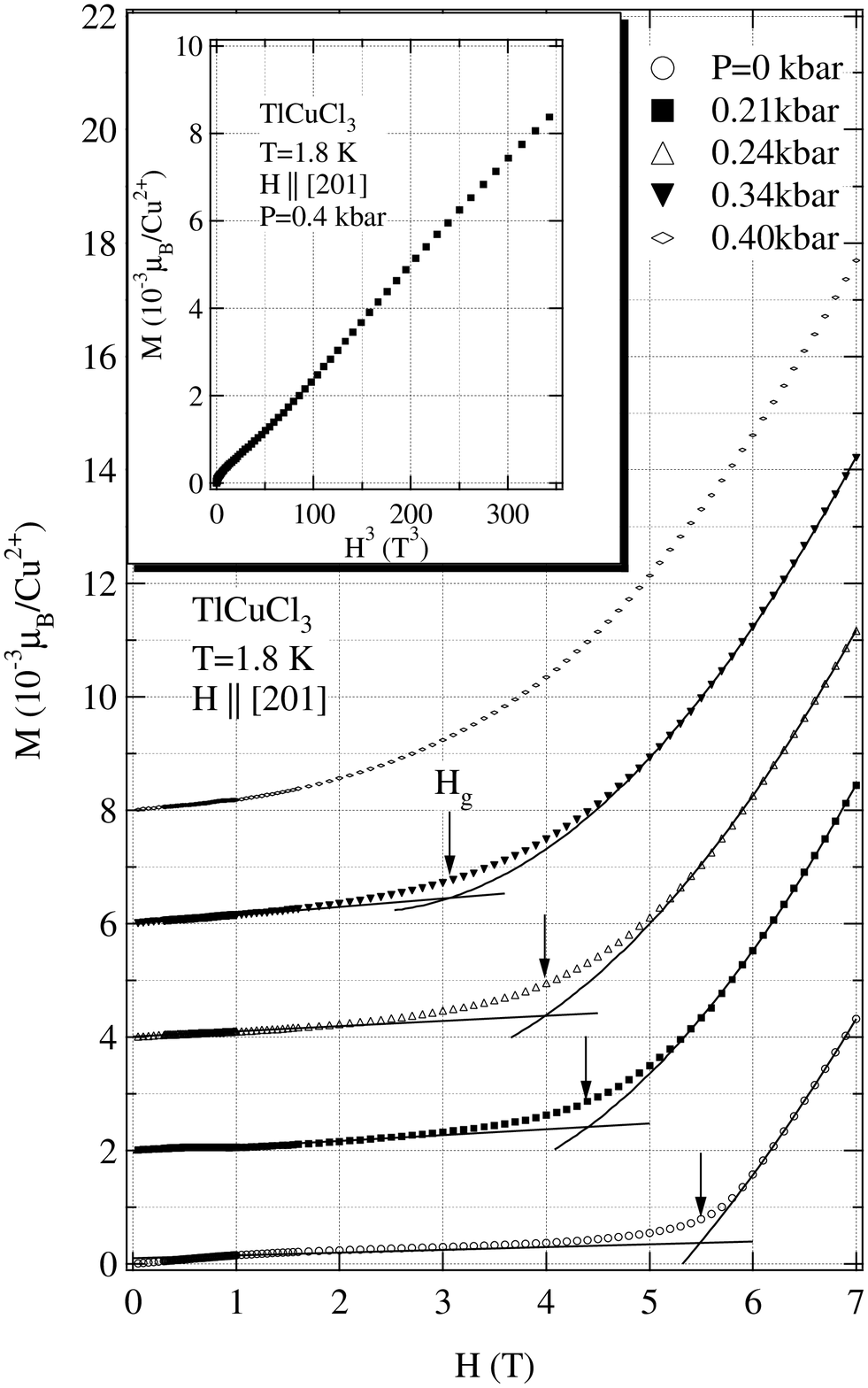}
\end{center}
\begin{center}
\includegraphics[width=7.8cm,clip]{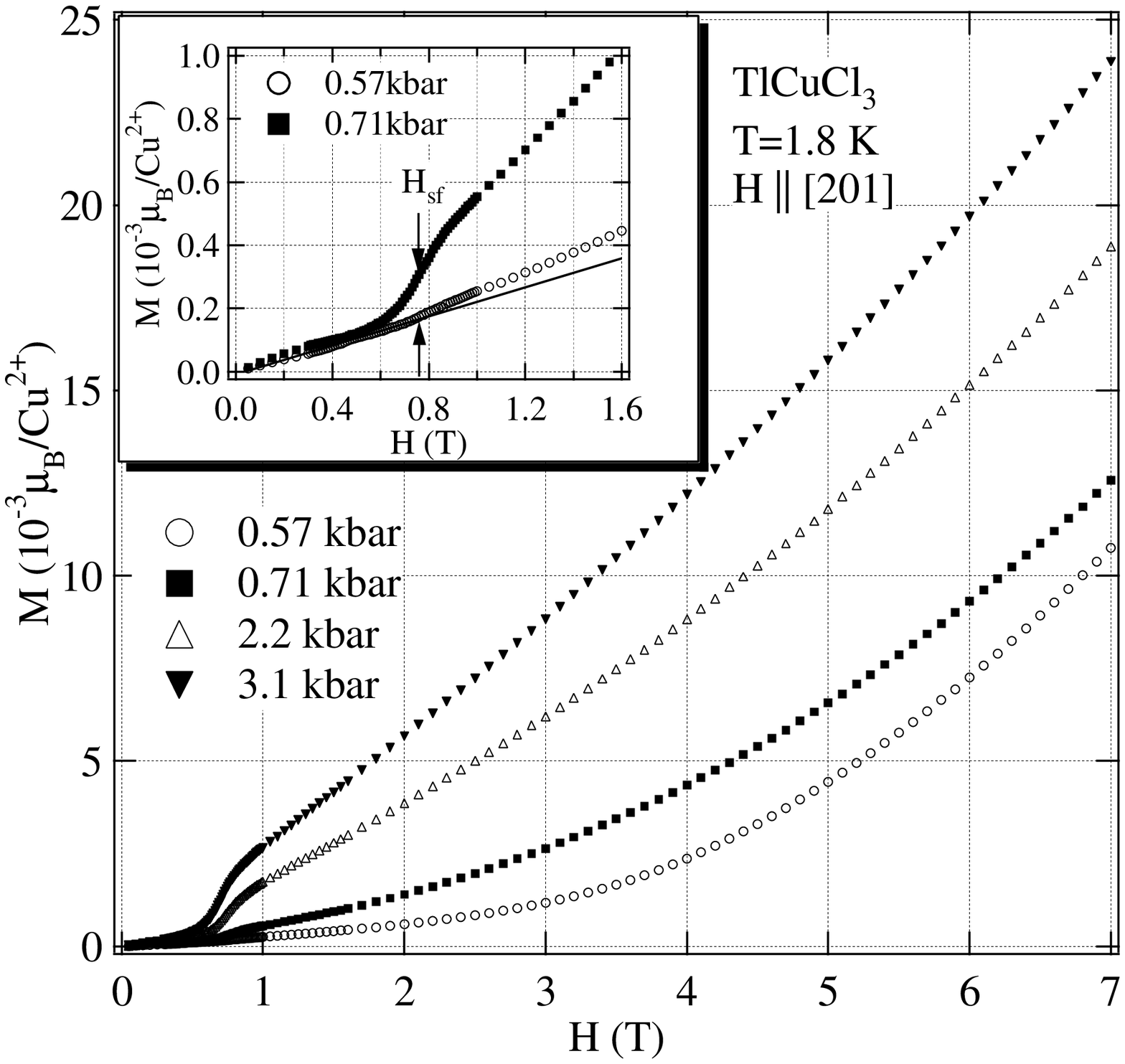}
\end{center}
\caption{Magnetization curves for $H\parallel [2, 0, 1]$ at $T=1.8$ K for (a) $P \leq 0.40$ kbar and (b) $P\geq 0.57$ kbar. Insets of (a) and (b) show magnetization vs $H^3$ at $P=0.40$ kbar and the enlargement of magnetization curves around the spin-flop field region for $P=0.57$ and 0.71 kbar, respectively. Solid lines in (a) denote two fitting functions used for $H < H_{\rm g}$ and $H > H_{\rm g}$.}
\label{fig:1}
\end{figure}

For $P \geq 0.57$ kbar, a spin-flop transition was observed at $H_{\rm sf}\simeq 0.7$ T, as shown in the inset of Fig. 1(b). This result indicates that the spin gap is already closed for $P \geq 0.57$ kbar, and that the ground state is an antiferromagnetic state with the easy-axis close to the $[2, 0, 1]$ direction.
For $P=0.44$ kbar, a tiny spin-flop transition was observed at 1.8 K, but not for $P=0.40$ kbar. Therefore, the critical pressure $P_{\rm c}$ at which the spin gap closes is between 0.40 and 0.44 kbar. In our previous short note \cite{Tanaka2}, we reported $P_{\rm c}\sim 2$ kbar. This value is incorrect. The error may have arisen from the use of a large applied magnetic field of 100 Oe for measuring the $T_{\rm c}$ of tin and the residual magnetic flux, which could not be removed completely. 

Figure 2 shows a plot of the critical field $H_{\rm g}$ as a function of pressure $P$. The gap ${\it \Delta}$ decreases with applied pressure, and closes at $P_{\rm c}\simeq 0.4$ kbar. The QPT at $P=P_{\rm c}$ seems to be of the second order, because the pressure dependence of the gap is expressed by the power law ${\it \Delta} \propto (P_{\rm c}-P)^{\alpha}$. The exponent ${\alpha}$ and critical pressure $P_{\rm c}$ obtained from the best fit are ${\alpha}=0.33\pm 0.05$ and $P_{\rm c}= 0.42\pm 0.05$ kbar.

In the dimer mean-field approximation, in which the triplet states are assumed to be created only on the dimer sites and the interdimer interactions are subjected to the mean-field approximation, the gap is expressed as 
$$
{\it \Delta}=\sqrt{J^2 - 2|{\tilde J}|J}, \eqno (1)
$$ 
where ${\tilde J}$ is expressed by a certain linear combination of interdimer interactions \cite{Oosawa4}. If the interdimer interactions have a linear pressure dependence, then we have ${\it \Delta} \propto (P_{\rm c}-P)^{1/2}$. The experimental value of $\alpha$ is approximately 2/3 times the theoretical value of $\alpha =1/2$. Since the decrease in the gap near $P_{\rm c}$ is rapid, there is also the possibility of a first order transition due to spin-lattice coupling as reported for the field-induced magnetic ordering in TlCuCl$_3$ \cite{Vyaselev}.
\begin{figure}
\begin{center}
\includegraphics[width=7.8cm,clip]{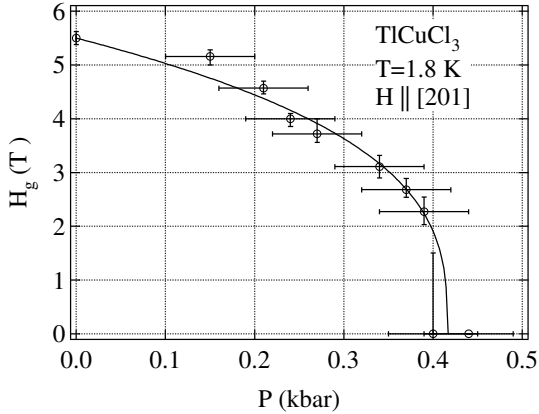}
\end{center}
\caption{Critical field $H_{\rm g}={\it \Delta}/g\mu_{\rm B}$ as a function of pressure $P$. Solid line is a fit using a power law with ${\alpha}=0.33$ and $P_{\rm c}= 0.42$ kbar.}
\label{fig:2}
\end{figure}

Oosawa {\it et al.} \cite{Oosawa5,Oosawa6} studied pressure-induced magnetic ordering in TlCuCl$_3$ at $P=14.8$ kbar in neutron scattering experiments. Below the ordering temperature $T_{\rm N}=16.9$ K, they observed magnetic Bragg reflections at ${\mib Q}=(h, 0, l)$ with the integer $h$ and the odd integer $l$, which are equivalent to those for the lowest magnetic excitation at zero pressure. The results of the present work and those of Oosawa {\it et al.} clearly indicate that the spin gap in TlCuCl$_3$ closes due to the applied pressure, and that the QCP is realized at $P_{\rm c}=0.42$ kbar and $T=0$. When interdimer exchange interactions are enhanced, the bandwidth of magnon dispersion is enhanced, thus the spin gap is reduced, as shown in eq. (1). Therefore, we can deduce that the pressure-induced QPT in TlCuCl$_3$ is caused by the relative enhancement of the interdimer exchange interactions against the intradimer exchange interaction due to the applied pressure. 

Pressure-induced magnetic ordering in the spin gap system was reported for Cu$_2$(C$_5$H$_{12}$N$_2$)$_2$Cl$_4$ \cite{Mito}. The ordering mechanism was interpreted as follows. Parts of singlet spin pairs are broken by the local deformation of the crystal lattice, which is evident in the Curie term of the paramagnetic susceptibility enhanced by the applied pressure. The induced magnetic moments around the broken spin pairs interact through effective exchange interactions mediated by intermediate singlet spins and form long range ordering similar to the impurity-induced ordering in the spin gap system \cite{Hase,Regnault}. The singlet-triplet excitation gap remains even in the ordered state. Therefore, the mechanisms leading to the pressure-induced magnetic ordering in Cu$_2$(C$_5$H$_{12}$N$_2$)$_2$Cl$_4$ and TlCuCl$_3$ are different. To the best of the authors' knowledge, TlCuCl$_3$ shows the first example of a pressure-induced magnetic QPT due to the closing of the spin gap.

As shown in Fig. 1(a), the magnetization curve for $P\approx P_{\rm c}$ is largely rounded, as if the spin gap and magnetic ordering coexist. Matsumoto {\it et al.} \cite{Matsumoto2} argued theoretically that the low-field magnetization at $P=P_{\rm c}$ is proportional to $H^3$, and thus, different from the usual magnetization curve, and showed that the magnetization per dimer $m$ is given by
$$
m\simeq \left(g\mu_{\rm B}H/J\right)^3, \eqno (2)
$$
in units of $g\mu_{\rm B}$. This is because the two triplet components $\left|1,1\right>$ and $\left|1,-1\right>$ contribute equally to the ground state at zero field. In the inset of Fig. 1, we plot magnetizations at $P=0.40$ kbar being close to $P_{\rm c}=0.42$ kbar as a function of $H^3$. We can see that the magnetization for $P\approx P_{\rm c}$ is approximately proportional to $H^3$, as theoretically predicted. However, the intradimer exchange interaction $J/k_{\rm B}= 48$ K evaluated by applying eq. (2) to the experimental data shown in the inset of Fig. 1 is smaller than $J/k_{\rm B}=65.9$ K obtained from dispersion relations at ambient pressure \cite{Cavadini1,Oosawa2}. Since the critical pressure $P_{\rm c}= 0.42$ kbar is small, the values of $J$ at $P_{\rm c}$ and ambient pressure should almost be the same. The discrepancy between the two $J$ values may arise from the renormalization effect in the magnetic excitations or from the quantum fluctuations corresponding to the increase in magnetization.

As shown in Fig. 1(b), the spin-flop transition is clearly observed for $H\parallel [2, 0, 1]$. However, the $[2, 0, 1]$ direction is not the easy-axis, because the magnetization curve for $H < H_{\rm sf}$ has a finite slope. In the impurity-induced antiferromagnetic state in Tl(Cu$_{1-x}$Mg$_{x}$)Cl$_3$, the spin easy-axis lies in the $a-c$ plane and is inclined from the $[2, 0, 1]$ direction by an angle of $13^{\circ}$ toward the $a$-axis \cite{Oosawa8}. The second easy-axis is the crystallographic $b$-axis, and the hard-axis lies in the $a-c$ plane. These magnetic principal axes were determined from the antiferromagnetic resonance modes \cite{Shindo2}. Since the anisotropy energy in pure and doped TlCuCl$_3$ arises from the anisotropic exchange interaction or the dipolar interaction, the easy-axes for pure and doped TlCuCl$_3$ should be close to each other. It is well-known that for a biaxial antiferromagnet when the external field is inclined from the easy-axis to the hard-axis, the spin-flop transition occurs as long as the easy-axis component of the external field is outside the so-called critical hyperbola, while when the external field is inclined from the easy-axis to the second easy-axis, the spin-flop transition disappears rapidly. From these reasons, we can deduce that also in the pressure-induced ordered state of TlCuCl$_3$, the easy-axis lies in the $a-c$ plane and is inclined from the $[2, 0, 1]$ direction by an angle of $\sim 10^{\circ}$ toward the $a$-axis, and that the $b$-axis is the second easy-axis. 

In the classical spin model, the spin-flop transition field $H_{\rm sf}$ is proportional to the square root of the product of the anisotropy and exchange fields, both of which are proportional to the average spin moment $\langle S\rangle$. Since $\langle S\rangle$ increases with increasing pressure for $P > P_{\rm c}$, the spin-flop field $H_{\rm sf}$ should increase with applied pressure, if the anisotropy and exchange constants remain unchanged. However, as shown in Fig. 1(b), the spin-flop field $H_{\rm sf}$ does not significantly depend on pressure. We discuss this problem according to the dimer mean-field theory \cite{Oosawa4}. We assume that the magnetic anisotropy arises from the anisotropy of the intradimer exchange interaction. Taking the easy-axis and the second easy-axis as the $z$- and $x$-directions, respectively, we write the difference in energy between the cases where spins point the $z$- and $x$- directions as ${\it \Delta}JS_1^zS_2^z$. We treat the anisotropy energy by the mean-field approximation. The basis state of a dimer is expressed as 
$
\psi=\left|0,0\right> \cos{\theta}+\left(\left|1,1\right> \cos{\varphi} - \left|1,-1\right> \sin{\varphi} \right)\sin{\theta}, 
$
where angles $\theta$ and $\varphi$ are introduced to satisfy the normalization condition, and the phases of triplets are omitted for simplification. Using the above basis state, we calculate the average values of spin operators (for details, see Appendix of ref. 15). For the present pressure range and $H \leq H_{\rm sf}\simeq 0.7$ T, $\sin^2\theta \ll 1$ and $\cos 2\varphi \ll 1$, {\it e.g.}, $\sin^2{\theta}=0.13$ at $P=14.8$ kbar \cite{Oosawa6}. In the antiferromagnetic state, ${\varphi}={\pi}/4$, $\cos 2{\theta}=J/(2|{\tilde J}|)$ and $\left<S^{z}_{1} \right>=-\left<S^{z}_{2}\right>=-\cos{\theta}\sin{\theta}$. The energy is independent of $H$.
On the other hand, in the spin-flop state, $\theta$, $\varphi$ and the average values of spin operators are expressed as shown in the Appendix of ref. 15. 
Comparing the energies of the antiferromagnetic state and the spin-flop state, we obtain 
$
g\mu_{\rm B}H_{\rm sf}\simeq \sqrt{4{\it \Delta}JJ/3}. 
$
Here, we used the relation $J\simeq 2|{\tilde J}|$ in the present pressure range. The spin-flop field $H_{\rm sf}$ has no singularity at $P_{\rm c}$. Since $J$ and ${\it \Delta}J$ should not markedly change with pressure, $H_{\rm sf}$ is almost independent of pressure. This is consistent with experimental results. Substituting $J/k_{\rm B}=65.9$ K, $H_{\rm sf}=0.7$ T and $g=2.06$, as determined by ESR, we obtain ${\it \Delta}J/k_{\rm B}=1\times10^{-2}$ K. This indicates that the anisotropy in the plane, which is perpendicular to the hard-axis is significantly small.

For $P > 2.0$ kbar, the magnetic susceptibility $M/H$ measured at $H=0.1$ T exhibits a clear bend anomaly due to magnetic ordering as shown in the inset of Fig. 3, while for $P < 2.0$ kbar, magnetization exhibits a small anomaly at the ordering temperature $T_{\rm N}$ for the magnetic field $H < H_{\rm sf}$. Thus, $T_{\rm N}$ for $P < 2.0$ kbar was measured at $H=1.0$ T, which is higher than $H_{\rm sf}$. This should be appropriate, because the values of $T_{\rm N}$ at $H=0.1$ and 1.0 T are almost the same for $P > 2.0$ kbar, as shown in the inset of Fig. 3. For $H=1.0$ T, a bend anomaly is clearly observed at $T_{\rm N}$, because the susceptibility for $H > H_{\rm g}$ is larger than that for $H < H_{\rm g}$. Figure 3 shows a plot of $T_{\rm N}$ as a function of applied pressure. The N\'{e}el temperature  $T_{\rm N}$ increases with increasing pressure. This is because the sublattice magnetization at zero temperature is enhanced with applied pressure. The pressure dependence of $T_{\rm N}$ can be described by the power law $T_{\rm N} \propto (P-P_{\rm c})^{\beta}$. The best fit is obtained with $\beta=0.44\pm 0.05$. 
\begin{figure}
\begin{center}
\includegraphics[width=7.8cm,clip]{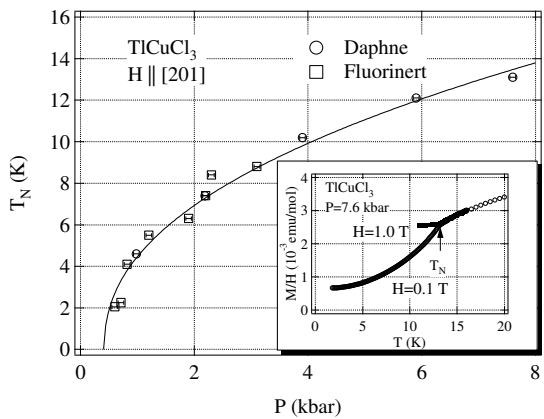}
\end{center}
\caption{Plot of N\'{e}el temperature $T_{\rm N}$ as a function of pressure $P$. Solid line is a fit using a power law with $\beta=0.44$. Inset shows the low-temperature susceptibilities $M/H$ for $P=7.6$ kbar. Arrow indicates the N\'{e}el temperature.}
\label{fig:3}
\end{figure}

In conclusion, we have presented the results of magnetization measurements on the spin gap system TlCuCl$_3$ under hydrostatic pressure for magnetic field $H$ parallel to the $[2, 0, 1]$ direction. A pressure-induced QPT from the gapped state to the antiferromagnetic state occurs at the critical pressure $P_{\rm c} = 0.42\pm 0.05$ kbar. The pressure dependence of the gap is described by the power law ${\it \Delta} \propto (P_{\rm c}-P)^{\alpha}$ with ${\alpha}=0.33\pm 0.05$. It is deduced that the applied pressure enhances the interdimer exchange interactions relative to the intradimer exchange interaction, thus the small spin gap shrinks and closes. The magnetization at $P \approx P_{\rm c}$ is approximately proportional to $H^3$, as theoretically predicted \cite{Matsumoto2}. For $P > P_{\rm c}$, a spin-flop transition was observed at $H_{\rm sf}\sim 0.7$ T. The spin-flop field does not exhibit any singularity at $P_{\rm c}$. The N\'{e}el temperature increases with pressure. The pressure dependence of $T_{\rm N}$ is described by the power law $T_{\rm N} \propto (P-P_{\rm c})^{\beta}$ with $\beta=0.44\pm 0.05$.

The authors would like to thank A. Oosawa and M. Matsumoto for stimulating discussions. This work was supported by the Toray Science Foundation, and a Grant-in-Aid for Scientific Research on Priority Areas and a 21st Century COE Program at Tokyo Tech ``Nanometer-Scale Quantum Physics'' both from the Ministry of Education, Culture, Sports, Science and Technology of Japan.

\end{document}